\documentclass{jps-cp}
\usepackage{txfonts} 
\usepackage{graphicx}
\usepackage{subfigure}

\title{Strangeon Matter in a Liquid Drop Model}

\author{Zheng \textsc{Wang}$^{1,2}$, Jiguang \textsc{Lu}$^{3,2}$ and Renxin \textsc{Xu}$^{1,2,4}$}

\inst{$^{1}$State Key Laboratory of Nuclear Science \& Technology, Peking University, Beijing 100871, China\\
$^{2}$Department of Astronomy, School of Physics, Peking University, Beijing 100871, China\\
$^{3}$National Astronomical Observatories, Chinese Academy of Sciences, Beijing 100012, China\\
$^{4}$Kavli Institute for Astronomy and Astrophysics, Peking University, Beijing 100871, China}

\email{zhengwang@pku.edu.cn, lujig@pku.edu.cn, r.x.xu@pku.edu.cn}

\recdate{May 31, 2017}

\abst{%
The liquid drop model of 2-flavored ($u$ and $d$) nucleus is well known and successful, analogically, a similar drop model for 3-flavored ($u$, $d$ and $s$) nucleus is developed.
A 3-flavored nucleus conjectured could be stable only if its baryon number is larger than a critical one, $A_{\rm c}$, in which strangeons are the constituent as an analogy of nucleons for nucleus~\cite{lx17}.
We try to model strangeon matter in a sense of phenomenological liquid drop, with two free parameters: the mass per baryon of a strangeon in vacuum, $M$, and potential depth between strangeons, $\epsilon$.
It is found that, for $M\sim$ GeV and $\epsilon\sim 100$ MeV, strangeon matter could be stable and its critical number could be as low as $A_{\rm c}=300$.
}%
\kword{pulsar: general, dense matter, equation of state}

\begin{document}
\maketitle

The Witten's conjecture~\cite{witten84} about dense matter could be extended to a generalized version: strange matter in bulk could be absolutely stable, in which quarks are either free (for strange quark matter) or localized (for strangeon matter)~\cite{lx17,X03}.
A strangeon may contain equal numbers of $u$, $d$ and $s$ quarks, and its number of quarks inside could be 6, 9, 12, 18 or even more.
The interaction between strangeons could be similar to that between nucleons (i.e., Lennard-Jones-like), and consequently, it might then be rational to have a description of strangeon matter based on this liquid drop scenario, a model that has already been certified for nucleus and is always introduced in standard text books.

Strangeon matter, formerly known as solid quark matter~\cite{X03}, is conjectured when chiral
symmetry is broken in the QCD phase-diagram.
However, a quarkyonic phase is suggested when chiral symmetry is restored but quarks are confined~\cite{M07}.
The color interaction between quarks in the strangeon phase might be stronger than that inside quarkyonic matter, but the details yet remains unclear.

\section{A Model of Strangeon Matter}

In the conventional liquid drop model of nucleus, there are terms of volume, surface, Coulomb, and others~\cite{K03},
\begin{equation}
E(A,Z)=ZM_\mathrm{p}+(A-Z)M_\mathrm{n}+b_\mathrm{vol} (1-k_\mathrm{vol}
I^2)A+b_\mathrm{surf} (1-k_\mathrm{surf} I^2)A^{2/3}+\frac{3}{5}\frac{e^2Z^2}{r_0A^{1/3}},
\label{eq1}
\end{equation}
where $A$ and $Z$ are the baryon number and proton number of the nucleus; $I=(N-Z)/A$, represents the asymmetry of the nucleus; $M_\mathrm{p}$ and $M_\mathrm{n}$ are the mass of proton and neutron; $b_\mathrm{vol}$,
$k_\mathrm{vol}$, $b_\mathrm{surf}$, $k_\mathrm{surf}$ and $r_0$ are the
parameters of the model.

In analogy to the drop model of nucleus, a liquid drop model could be constructed for strangeon matter.
The asymmetry energy, however, could be negligible due to three-flavor symmetry restoration, whereas the volume and surface terms remain, and the energy per baryon of a strangeon drop could then read (the Coulomb energy is also not significant),
\begin{equation}
E/A=M+b_\mathrm{vol,~s}+b_\mathrm{surf,~s} A^{-1/3},
\label{eq2}
\end{equation}
where $M$ is the mass per baryon of a strangeon in vacuum,
and the subscript ``s'' denotes parameters for strangeon drop.
Certainly a strangeon could not be stable in vacuum due to both strong and weak interactions, and it seems that parameters of $M$ and $b_\mathrm{vol,~s}$ are degenerated. However, note that $b_\mathrm{vol}$ and $b_\mathrm{surf}$ are related in Eq.~(\ref{eq1}). A similar relation between $b_\mathrm{vol,~s}$ and $b_\mathrm{surf,~s}$ will break the degenerate.

One may expect that strangeon matter could be more stable than 2-flavored nucleus when its baryon number, $A$, is greater than a critical baryon number, $A_{\rm c}$, if the general Witten's conjecture is correct.
It is evident from Eq.~(\ref{eq2}) that the energy per baryon, $E/A$, decreases as the baryon number $A$ increases for strangeon matter (note that $b_\mathrm{surf,~s}$ is always positive), while that energy is minimum for $^{56}$Fe but increases for nuclei if $A>56$.
The $A_{\rm c}$-number is determined by equating these two energies.
Certainly, in order to obtain the $A_{\rm c}$-number, we should have the parameter of $b_\mathrm{vol,~s}$, which are focused in the next section.
In addition, for the $b_\mathrm{surf,~s}$-parameter, we make use of a same scale of 2-flavored drop model in which one has approximately $b_\mathrm{surf}\propto b_\mathrm{vol}$.

\section{A Corresponding State Approach and Results}

The law of corresponding states is proposed in a dimensionless way, which shows that the equation of state of substances with same form of interaction can be written in a reduced and universal form.
With this corresponding state approach, the parameters for strangeon
matter could be derived.

For the sake of simplicity, the interaction between two strangeons is assumed to be Lennard-Jones-like~\cite{lai09}, which is similar to the interaction
between atom of inert gas,
\begin{equation}
\phi(r)=\epsilon \left\{\frac{4}{(r/\sigma)^{12}} -\frac{4}{(r/\sigma)^6} \right\},
\end{equation}
where $\epsilon$, $\sigma$ are characteristic energy (i.e., potential depth) and length-scale of the interaction, respectively.
Based on the law of corresponding states, the interaction energy (thus $b_\mathrm{vol,~s}$ and $b_\mathrm{surf,~s}$) of
strangeon in a drop could be obtained~\cite{guo14}.
With these two parameters, the energy per baryon of a strangeon drop could be calculated with Eq~(\ref{eq2}).

\begin{figure}[tbh]
\centering
\subfigure{
\includegraphics[width=0.46\textwidth]{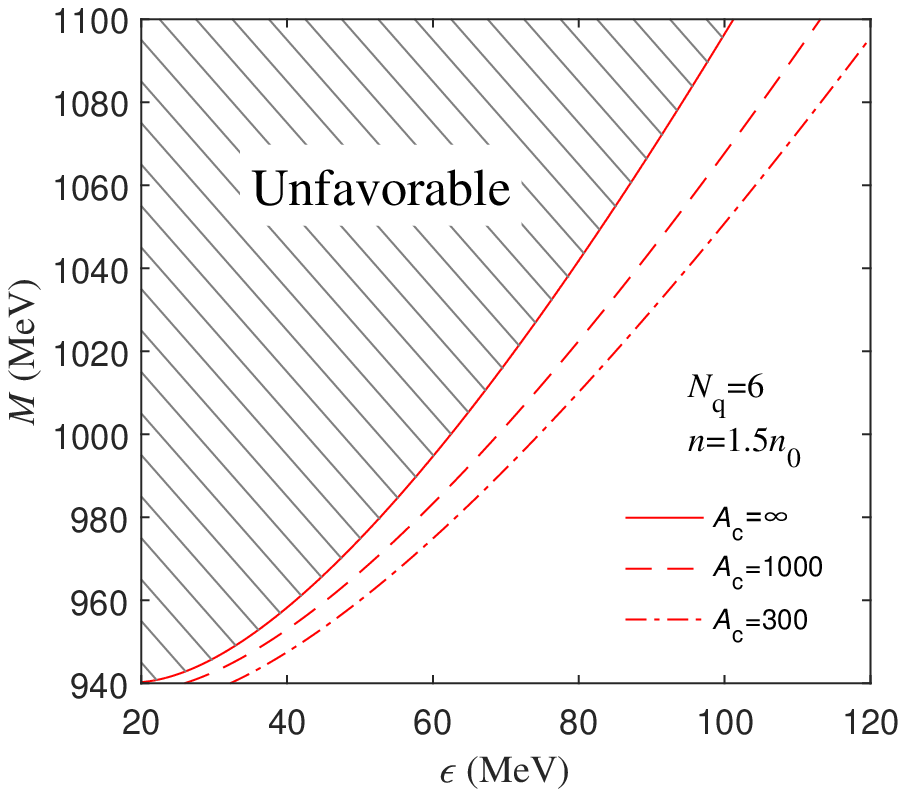}
\includegraphics[width=0.46\textwidth]{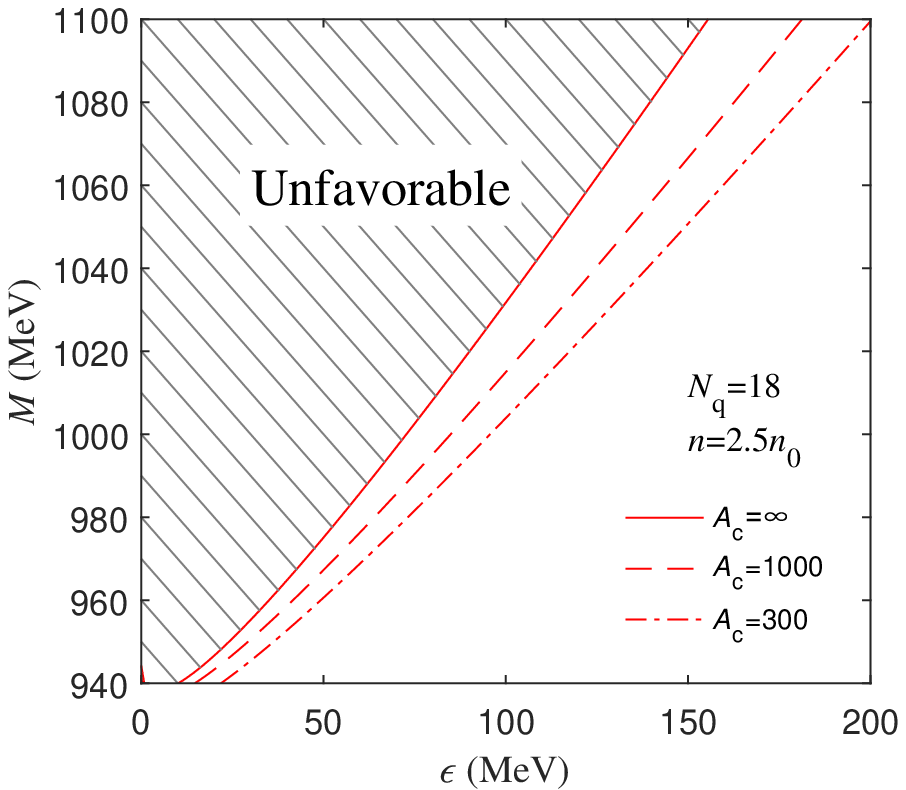}
}
\caption{The parameter space of stable strangeon matter in $M-\epsilon$ diagram, where $M$ is the mass per baryon of a strangeon in vacuum and $\epsilon$ is the potential depth.
We calculate with different parameter sets in a liquid drop model of strangeon matter, here is two of them: \{$N_{\rm q} = 6, n=1.5n_0$\} (left plot) and \{$N_{\rm q} = 18, n=2.5n_0$\} (right plot), where $N_{\rm q}$ is quark number of a strangeon and $n$ is the baryon number density.
It is evident that strangeon matter could be absolutely stable in a huge space of parameters.
}
\label{f1}
\end{figure}

For a given baryon number density, $n$ (to be a few nuclear density, $n_0$), and the quark number of one strangeon, $N_\mathrm{q}$, we have two free parameters to calculate the energy per baryon ($E/A$): $M$ and $\epsilon$.
If strangeon matter exits stably, the energy per baryon of strangeon drop should be lower than that of normal nuclei with baryon number $A>A_\textrm{c}$.
The $A_{\rm c}$-number as a function of those two free parameters $M$ and $\epsilon$ is shown in Fig.~\ref{f1}, for two parametric sets of \{$N_{\rm q} = 6, n=1.5n_0$\} and \{$N_{\rm q} = 18, n=2.5n_0$\}, respectively.
Even the surface energy is not included, the energy per baryon of strangeon matter cannot be lower than that of 2-flavored nucleus ($\sim 938$ MeV) in the hatched region labelled ``Unfavorable''.
As shown in the figure, there is huge parameter space where 3-flavored strangeon matter could be more stable than 2-flavored nucleon matter (i.e., nucleus).
It is worth noting that some of normal nuclei would be metastable (but strangeon matter is absolutely stable) if $A_{\rm c}\sim 100$ or even smaller.

\section{Discussions and Conclusions}

%

Conventionally, we call our model ``liquid drop model'', but the real strangeon matter could be ``solid'' in fact.
A strangeon is much more massive than a nucleon, and its quantum effect could then be weaker and its wave packet could be even smaller than the separation between strangeons.
Therefore, when the temperature of strangeon matter is significantly lower than
$\sim 0.1\epsilon\sim 10$ MeV, a strangeon drop should be solidified though its state equation may not change significantly.

%
%

There could be three different manifestations of strangeon matter~\cite{xu15}: strangeon star, strangeon cosmic ray, and even strangeon dark matter.
Normal 2-flavored baryonic matter would be extremely compressed by self-gravity of a core inside an evolved massive star. This compressed baryonic matter could finally be converted to strangeon matter via weak and strong interactions, and a pulsar-like compact star (i.e., strangeon star) with mass of $\sim M_\odot$ is created.
Binary merger of strangeon star might also eject relativistic strangeon nuggets detectable as cosmic rays, and such a strangeon cosmic ray
might be identified in an air-shower event.
Additionally, it is also possible to synthesize strangeon matter during the cosmic separation of QCD phases, and this kind of strangeon matter could have survived and exist in the present Universe, to be manifested in the form of dark matter.

In summary, we have tried to model strangeon matter in a sense
of phenomenological liquid drop, with the help of the law of corresponding states to derive model parameters.
It is found that strangeon matter could be stable even its baryon number to be as low as $300$ if $M\sim$ GeV and $\epsilon\sim 100$ MeV.
There is huge parameter space for strangeon matter to exist stably in this liquid drop model.
Interestingly, it supplies a unique possibility to explain very different manifestations in the Universe (the nature of pulsar, cosmic ray, and dark matter) with the strangeon matter conjecture.

\vspace{0.6cm}
\noindent
{\bf Acknowledgments:}
We are grateful to the members at the pulsar group of Peking University.
This work is supported by the National Natural Science Foundation of China (11673002 and U1531243) and the Strategic Priority Research Program of CAS (No. XDB23010200).

\end{document}